%% file: main.tex
\documentclass[sigconf,screen,nonacm]{acmart}

\usepackage[skins]{tcolorbox}
\usepackage{afterpage}

\usepackage{float}
\usepackage{microtype} 
\usepackage{graphicx} 

\usepackage{listings}
\usepackage{xcolor}
\usepackage{booktabs}
\usepackage{multirow}
\usepackage{fvextra}
\usepackage{fancyvrb}
\usepackage{adjustbox}
\usepackage{footnote}
\usepackage{fontawesome5}

\renewcommand\footnotetextcopyrightpermission[1]{}   
\setcopyright{none}   

\makesavenoteenv{tabular}
\DefineVerbatimEnvironment{HighlightVerbatim}{Verbatim}
{
    commandchars=\\\{\}, 
}

\definecolor{jsonkeys}{RGB}{17,99,216}
\definecolor{jsonstring}{RGB}{0,128,0}
\definecolor{jsonnumber}{RGB}{102,102,102}
\definecolor{jsonbrace}{RGB}{150,0,0}

\lstdefinestyle{json}{
  language=JSON,
  basicstyle=\footnotesize\ttfamily,
  numbers=none,
  numberstyle=\tiny,
  stepnumber=1,
  numbersep=8pt,
  showstringspaces=false,
  breaklines=true,
  frame=none,
  backgroundcolor=\color{white},
  literate=
    *{0}{{{\color{jsonnumber}0}}}{1}
     {1}{{{\color{jsonnumber}1}}}{1}
     {2}{{{\color{jsonnumber}2}}}{1}
     {3}{{{\color{jsonnumber}3}}}{1}
     {4}{{{\color{jsonnumber}4}}}{1}
     {5}{{{\color{jsonnumber}5}}}{1}
     {6}{{{\color{jsonnumber}6}}}{1}
     {7}{{{\color{jsonnumber}7}}}{1}
     {8}{{{\color{jsonnumber}8}}}{1}
     {9}{{{\color{jsonnumber}9}}}{1}
     {:}{{{\color{jsonbrace}{:}}}}{1}
     {,}{{{\color{jsonbrace}{,}}}}{1}
     {\{}{{{\color{jsonbrace}{\{}}}}{1}
     {\}}{{{\color{jsonbrace}{\}}}}}{1}
     {[}{{{\color{jsonbrace}{[}}}}{1}
     {]}{{{\color{jsonbrace}{]}}}}{1},
  sensitive=true,
  morestring=[b]",
  stringstyle=\color{jsonstring},
  keywordstyle=\color{jsonkeys},
  commentstyle=\color{jsonnumber}\ttfamily,
}

\usepackage{hyperref}
\newcommand\myshade{85}
\definecolor{mylinkcolor}{RGB}{0,0,139}      
\definecolor{mycitecolor}{RGB}{128, 18, 10}  
\definecolor{myurlcolor}{RGB}{67,47,136}        
\usepackage{makecell}
\hypersetup{
  linkcolor  = mylinkcolor!\myshade!black,
  citecolor  = mycitecolor!\myshade!black,
  urlcolor   = myurlcolor!\myshade!black,
  colorlinks = true,
}

\usepackage{todonotes} 
\usepackage{xargs} 
 
\newcommand{\method}{\textsc{BitsAI-CR}}

\newcommandx{\info}[2][1=]{\todo[linecolor=red,backgroundcolor=red!25,bordercolor=red,#1]{#2}}
\newcommand{\find}[1]{
	\begin{tcolorbox}[tile,size=fbox,boxsep=1mm,boxrule=0pt,top=0pt,bottom=0pt,
		borderline west={0.2mm}{0pt}{black!50!white},colback=black!5!white]
		#1
	\end{tcolorbox}
}

\AtBeginDocument{%
  }

\begin{document}

\title{\method{}: Automated Code Review via LLM in Practice}
\author{
Tao Sun, Jian Xu$^{\dagger}$, Yuanpeng Li, Zhao Yan, Ge Zhang, 
Lintao Xie, Lu Geng, Zheng Wang, \\ Yueyan Chen, 
Qin Lin, Wenbo Duan, Kaixin Sui
}
\affiliation{\country{} \institution{ByteDance}}

\thanks{$^\dagger$ Corresponding Author.}

\renewcommand{\shortauthors}{Tao Sun et al.}

\begin{abstract}
Code review remains a critical yet resource-intensive process in software development, particularly challenging in large-scale industrial environments. While Large Language Models (LLMs) show promise for automating code review, existing solutions face significant limitations in precision and practicality. 
This paper presents \method{}, an innovative framework that enhances code review through a two-stage approach combining RuleChecker for initial issue detection and ReviewFilter for precision verification. The system is built upon a comprehensive taxonomy of review rules and implements a data flywheel mechanism that enables continuous performance improvement through structured feedback and evaluation metrics. Our approach introduces an Outdated Rate metric that can reflect developers' actual adoption of review comments, enabling automated evaluation and systematic optimization at scale. 
Empirical evaluation demonstrates \method{}'s effectiveness, achieving 75.0\% precision in review comment generation. For the Go language which has predominant usage at ByteDance, we maintain an Outdated Rate of 26.7\%. The system has been successfully deployed at ByteDance, serving over 12,000 Weekly Active Users (WAU). 
Our work provides valuable insights into the practical application of automated code review and offers a blueprint for organizations seeking to implement automated code reviews at scale.

\end{abstract}



\maketitle

\input{sections/01intro.tex}

\input{sections/02background.tex}

\input{sections/03method.tex}

\input{sections/04Experiments}

\input{sections/05Results}
\input{sections/06RelatedWork}

\section{Conclusion and Feature Work}

This paper presents \method{}, a comprehensive automated code review system that tackles both the efficiency bottlenecks in enterprise-scale review processes and the fundamental limitations of existing LLM-based solutions, particularly their insufficient comment effectiveness and lack of systematic improvement mechanisms. 
To achieve this, we introduce a novel taxonomy of review rules that serve as the foundation for our two-stage approach: RuleChecker for initial issue detection and ReviewFilter for precision enhancement. We further propose the Outdated Rate metric to evaluate comment practicality and drive systematic improvements through a data flywheel mechanism. Our empirical evaluation demonstrates the effectiveness of this approach, achieving an acceptable precision rate in comment generation and a competitive Outdated Rate in Go language reviews. The successful deployment at ByteDance validates its scalability and practical value in enterprise-scale software development environments.

Our future work will focus on these key directions for technical and service enhancement: First, we plan to expand language coverage from our current support of five mainstream programming languages to comprehensive coverage of all programming languages. Second, we will enhance our review rules, which currently focus primarily on function-level understanding with limited contextual information, by developing effective cross-file review capabilities. Through these continuous efforts, we strive to enrich and strengthen our system to provide more comprehensive and in-depth code review services, ultimately helping developers improve both code quality and engineering efficiency.

\bibliographystyle{ACM-Reference-Format}
\bibliography{main}

\end{document}

%% file: sections/01intro.tex
\section{Introduction}

Code review is a critical process in software development that significantly impacts code quality and project success~\cite{yangSurveyModernCode2024,bacchelli2013expectations}. 
It identifies potential security vulnerabilities, logical errors, and performance bottlenecks while facilitating knowledge sharing among R\&D teams\cite{sadowski2018modern}. 
However, our analysis at ByteDance reveals significant challenges in enterprise-scale code review practice. 
The internal data shows that reviewers spend an average of 15 minutes per review for over 50\% of cases, with the remaining cases taking even longer. 
Only 30\% of reviews receive immediate attention, while over 67\% of engineers express a need for more effective tools.

This evidence indicates a pressing demand for effective automated code review tools, yet current approaches~\cite {emanuelsson2008comparative,singh2017evaluating} remain inadequate or lack their practical impact comprehensive evaluation at an enterprise scale. 
Traditional static analysis tools, while offering basic code quality checks, introduce significant overhead to development workflows through complex build configurations and compilation. 
Large language models (LLMs) emerge as a promising solution due to their superior capabilities in code understanding and natural language generation~\cite{liu2024large,haider2024prompting,fanExploringCapabilitiesLLMs2024}. 
Recent commercial implementations have begun exploring LLM-based approaches to address these limitations, however, Google's approach~\cite{vijayvergiyaAIAssistedAssessmentCoding2024} only focuses on issue classification without generating specific review comments, while Tencent~\cite{yuFinetuningLargeLanguage2024} primarily addresses code maintainability concerns. 
Our investigation reveals three fundamental challenges in current LLM-based solutions: 
\textit{i)} insufficient precision in generating technically accurate comments, 
\textit{ii)} low practicality of comments that are technically correct but fail to provide substantial value~\cite{majumdar2022automated, naikCRScoreGroundingAutomated2024}, and 
\textit{iii)} lack of systematic mechanisms for targeted improvement, preventing data-driven evolution in both model precision and suggestion practicality.

To address these limitations, we present \method{}, a framework comprised of two primary components operating in synergy: a high-precision review comment generation pipeline and a data flywheel mechanism that continuously optimizes based on user feedback. 
First, to tackle the issue of low precision in review comments, the review generation pipeline is designed as a two-stage process called the RuleChecker and the ReviewFilter: 
(1) RuleChecker is a fine-tuned LLM  trained on a comprehensive taxonomy of 219 review rules, aimed at identifying potential issues, 
(2) ReviewFilter is another fine-tuned LLM followed by RuleChecker which improves precision by validating the detected issues. 
Second, to enable targeted continuous improvement, we construct a data flywheel that leverages real-world feedback for large-scale industrial scenarios: 
(1) leveraging the annotation feedback data to enhance training datasets, thereby improving the \method{}'s performance with targeted enhancements;
(2) metring the Outdated Rate to resolve the problem of technically correct but practically superfluous comments by automatically measuring the percentage of code lines modified after being flagged by \method{}, which provides concrete evidence of whether developers accept and act upon the review; 
and (3) dynamically adjusting review rules based on both Outdated Rates and precision measurements, removing rules that generate low-value comments with low Outdated Rates and high precision.
Through this approach, \method{} establishes a continuous improvement cycle to enhance code review.

Empirical evaluation shows that our \method{} takes significant improvements in precision, and developers are increasingly recognizing the generated comments. 
The two-stage review pipeline achieves a peak precision of 75.0\%, significantly reducing superfluous comments. 
More importantly, our data flywheel mechanism shows consistent improvement over time, with a 26.7\% Outdated Rate in Go. 
Large-scale industrial deployment at ByteDance further validates \method{}'s impact, engaging over 12,000 Weekly Active Users (WAU).

\begin{figure*}[t]
\centering
\includegraphics[width=0.98\linewidth]{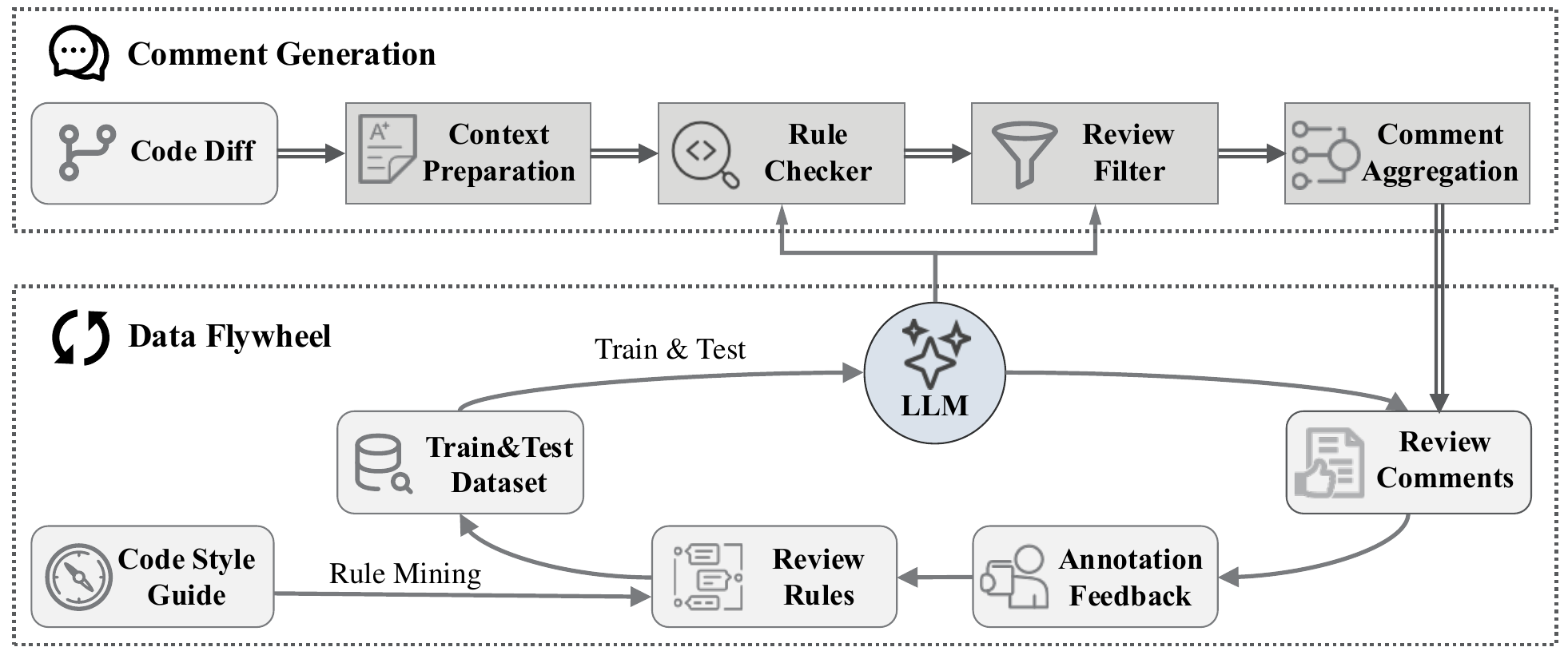}
\caption{The overview of \method{} framework for enhancing code review.}
\label{fig:cr}
\end{figure*}

This paper details how we build and deploy \method{}, used in the ByteDance production environment for several months and serving ten thousand engineers. The main contributions are:

\begin{itemize}
    \item A comprehensive taxonomy of review rules that streamlines the entire code review process, enabling evaluation practices, data collection mechanisms, and continuous optimization through a data-driven feedback loop.
    \item A two-stage approach combining RuleChecker for issue detection and ReviewFilter for verification, demonstrating substantial improvements in review effectiveness.
    \item The introduction of an Outdated Rate metric that addresses two key limitations of traditional precision measurements: the sustainability of manual evaluation and the ability to assess user acceptance of review comments.
    \item Empirical validation through large-scale deployment, demonstrating the data flywheel's effectiveness and scalability in real-world software development environments through systematic implementation practices.
\end{itemize}

%% file: sections/02background.tex
\section{Background}

Code review evolves from a simple correctness-checking mechanism to a comprehensive quality assurance process that shapes organizational culture and development practices. Modern reviews now evaluate multiple dimensions of code quality, including design patterns, architectural choices, and maintainability, while fostering knowledge sharing and team collaboration\cite{eldh2024code}. Meanwhile, code best practices\cite{mcintosh2016empirical} refers to a set of widely accepted methodologies that prove to enhance the quality and efficiency of the software development process. Code review serves as a crucial means to implement these best practices, with reviewers evaluating various aspects of the code based on the company's internal standards.

At ByteDance, code review is an integral part of the software development lifecycle, following industry-standard practices while incorporating company-specific requests. Similar to existing code management systems like GitHub's pull request\cite{kalliamvakou2014promises} and Google's Piper\cite{potvin2016google}, ByteDance's software development process requires experienced reviewers to examine every code change submission. The code is only merged after receiving approval for ``LGTM'' (i.e., ``looks good to me''). Authors and reviewers exchange opinions through the code review system, examining snapshots of affected files.

The typical incremental review process involves several steps: (1) A developer creates a Merge Request (MR). (2) The code diff is submitted to reviewers for error identification and resolution, which can be automated or enhanced by \method{}. (3) Reviewers, assisted by or replaced with \method{}, provide feedback to ensure the code meets quality standards. (4) The developer refactors the code based on reviewer feedback. (5) Steps 2-4 are repeated as necessary. (6) Once reviewers confirm all issues are resolved, the code receives approval for merging.

%% file: sections/03method.tex
\section{Methodology}

In this section, we present our approach to automated code review. We first introduce an overview of our framework that illustrates how the Review Comment Generation pipeline and Data Flywheel mechanisms work together as shown in Figure~\ref{fig:cr}. We then detail the taxonomy of code review rules that serve as the foundation for the entire system. Next, we elaborate on the review comment generation pipeline, which includes context preparation, rule checking, review filtering, and comment aggregation. Finally, we describe the data flywheel mechanism to continuously improve \method{}.

\subsection{Framework of \method{}}

\input{fig/prompt/task}

The task of \method{} is to analyze code changes and provide meaningful review comments. Given a code diff, the system aims to identify potential issues and generate appropriate review suggestions. The output includes the review category, specific problematic code locations, and detailed explanatory comments. Figure~\ref{fig:code-review-example} illustrates this input-output relationship. The input consists of code changes, while the output provides structured review feedback that helps developers improve their code quality.

As shown in Figure~\ref{fig:cr}, the \method{} framework consists of two primary components: a Review Comment Generation pipeline and a Data Flywheel mechanism for continuous improvement. The Review Comment Generation pipeline processes code reviews through four steps: (1) Context Preparation that structures the input for analysis, (2) RuleChecker that identifies potential issues using an LLM, (3) ReviewFilter that validates the detected issues, and (4) Comment Aggregation that consolidates similar feedback. 
Supporting this pipeline, the Data Flywheel mechanism ensures continuous improvement through the taxonomy of review rules, which are used to create training and testing datasets for the LLM. The generated review comments receive annotation feedback, which in turn helps refine the review rules, creating a self-improving system that enhances code review quality over time.

\subsection{The Taxonomy of Code Review Rules} \label{Taxonomy}

We establish a comprehensive taxonomy of code review rules which serves as a foundational component of our \method{} system, and introduce a three-tiered classification structure: review dimensions (broad areas of code quality assessment), review categories (specific issue types requiring evaluation) and review rules (detailed criteria with examples). The term ``review rules'' refers to detailed specifications that define concrete evaluation criteria. Each review category is backed by detailed rules that specify elements such as issue type classification, severity levels, comprehensive descriptions, and illustrative examples of both proper and improper implementations.
Our review dimensions encompass the unique characteristics of various programming languages, including language-specific idioms, performance optimizations, and common pitfalls associated with each language's ecosystem. As shown in Table~\ref{tab:code-review-taxonomy}, they include:
\begin{itemize}
    \item \textbf{Code Defect}: This dimension covers various aspects of code correctness, including error handling, logic flaws, and resource management.
    \item \textbf{Security Vulnerability}: It addresses critical security concerns such as injection vulnerabilities, cross-site scripting, and insecure data handling.
    \item \textbf{Maintainability and Readability}: It emphasizes code clarity, consistency, and structure to ensure the long-term maintainability of the codebase.
    \item \textbf{Performance Issue}: This category focuses on optimizing code execution, covering areas like efficient data structures, query optimization, and resource utilization.

\end{itemize}

\begin{table}[t!]
\caption{Distribution and Outdated Rates in the Taxonomy of Code Review Rules}
\label{tab:code-review-taxonomy}
\resizebox{\linewidth}{!}{
\begin{tabular}{clcc}
\toprule
\textbf{\begin{tabular}[c]{@{}c@{}}Review\\ Dimension\end{tabular}} &
  \multicolumn{1}{c}{\textbf{\begin{tabular}[c]{@{}c@{}}Review\\ Category\end{tabular}}} &
  \textbf{\begin{tabular}[c]{@{}c@{}}Distribution \\ (\%)\end{tabular}} &
  \textbf{\begin{tabular}[c]{@{}c@{}}Outdated Rate \\ (\%)\end{tabular}} \\ \midrule
\multirow{8}{*}{\textbf{\begin{tabular}[c]{@{}c@{}}Security\\ Vulnerability\end{tabular}}}           & SQL Injection                                 & 0.08  & 11.54 \\
                                                                                                     & Insecure Deserialization                      & 0.03  & 20.00 \\
                                                                                                     & Insecure Object Reference                     & 0.04  & 15.38 \\
                                                                                                     & Memory Leak - Long-term Reference Holding     & 0.05  & 37.50 \\
                                                                                                     & Improper Password Handling                    & 0.29  & 28.41 \\
                                                                                                     & Type and Non-null Assertion                   & 0.46  & 14.18 \\
                                                                                                     & Cross-Site Scripting (XSS)                    & 0.07  & 30.43 \\
                                                                                                     & Cross-Site Request Forgery (CSRF)             & 0.10  & 23.33 \\ \midrule
\multirow{11}{*}{\textbf{\begin{tabular}[c]{@{}c@{}}Code\\ Defect\end{tabular}}}                     & Function Parameter Passing                    & 0.71  & 27.73 \\
                                                                                                     & Loop Logic Errors                             & 0.39  & 33.61 \\
                                                                                                     & Database Access Error                         & 0.20  & 27.87 \\
                                                                                                     & Conditional Logic Error/Omission/Duplication  & 3.48  & 30.61 \\
                                                                                                     & Null Pointer Exception                        & 22.79 & 27.58 \\
                                                                                                     & Algorithm/Business Logic Error                & 3.50  & 18.89 \\
                                                                                                     & Index/Boundary Condition Error                & 3.57  & 26.73 \\
                                                                                                     & Syntax Issue                                  & 0.08  & 25.00 \\
                                                                                                     & Resource Not Released/Resource Leak           & 0.60  & 17.93 \\
                                                                                                     & Error and Exception Handling Issue            & 4.11  & 25.20 \\
                                                                                                     & Incorrect Concurrency Control                 & 0.65  & 17.91 \\
                                                                                                     & Data Format, Conversion, and Comparison Error & 1.30  & 41.37 \\
                                                                                                     & Incorrect Sequence Dependency                 & 0.03  & 0.00  \\ \midrule
\multirow{14}{*}{\textbf{\begin{tabular}[c]{@{}c@{}}Maintainability\\ and Readability\end{tabular}}} & Unclear Naming                                & 0.34  & 25.00 \\
                                                                                                     & Code Testability Issues                       & 0.05  & 26.67 \\
                                                                                                     & Code Readability                              & 5.00  & 15.82 \\
                                                                                                     & Code Formatting Errors/Inconsistencies        & 0.60  & 13.51 \\
                                                                                                     & Redundant/Complex Conditional Logic           & 2.24  & 20.99 \\
                                                                                                     & Variable Naming Conventions                   & 0.61  & 5.88  \\
                                                                                                     & Complex Code                                  & 0.80  & 23.48 \\
                                                                                                     & Spelling Error                                & 7.97  & 31.52 \\
                                                                                                     & Unused Definition/Redundant Code              & 1.83  & 21.63 \\
                                                                                                     & Missing or Inappropriate Code Comments        & 8.27  & 26.25 \\
                                                                                                     & Overly Long Functions or Methods              & 2.15  & 14.76 \\
                                                                                                     & Code Duplication                              & 7.23  & 17.88 \\
                                                                                                     & Unclear Error Handling                        & 0.11  & 17.14 \\
                                                                                                     & Magic Numbers/Strings                         & 20.62 & 18.87 \\ \midrule
\multirow{6}{*}{\textbf{\begin{tabular}[c]{@{}c@{}}Performance\\ Issue\end{tabular}}}                & Inappropriate Data Structures                 & 0.38  & 13.56 \\
                                                                                                     & Unoptimized Loops                             & 0.14  & 22.73 \\
                                                                                                     & Data Format Conversion Performance            & 0.37  & 29.57 \\
                                                                                                     & Excessive or Improper Lock Usage              & 0.01  & 0.00  \\
                                                                                                     & Excessive I/O Operations                      & 0.25  & 13.16 \\
                                                                                                     & Repeated Calculations                         & 0.01  & 0.00  \\ \bottomrule
\end{tabular}%
}
\end{table}

Each Review Category encompasses multiple specific review rules, enabling fine-grained quality assessment. For instance, consider the review dimension ``Code Defects'', which encompasses various review categories including ``Conditional Logic  Error/Omission/Duplication''. Within this category, specific review rules include ``unreachable code detection'' and ``infinite loop identification'', etc. These rules address critical code quality issues: ``unreachable code`` represents redundant statements that can never be executed due to logical constraints, while ``infinite loop identification`` identifies potentially endless program execution cycles that could lead to system resource exhaustion. We already include a range of programming languages including Go, JavaScript, TypeScript, Python, and Java, totalling 219 review rules.

This structure offers several key advantages: (1) Efficient Data Collection: The hierarchical taxonomy enables systematic data gathering and labeling, with clear guidelines for categorizing different types of code issues and their corresponding comments. (2) Streamlined Model Training: The structured categorization provides well-defined training objectives and organized datasets, leading to more effective model fine-tuning for specific review tasks. (3) Systematic Evaluation: The clear classification structure enables precise measurement of model performance across different review dimensions, facilitating comprehensive quality assessment. (4) Targeted Optimization: The structured feedback mechanism allows for systematic improvement of model performance through clearly defined enhancement paths for each review dimension.

\subsection{The Pipeline of Review Comment Generation}
Our review comment generation pipeline implements a two-stage approach centred on RuleChecker and ReviewFilter. The first stage employs RuleChecker to identify potential issues based on the taxonomy, while the second stage utilizes ReviewFilter to validate these findings, ensuring high precision. Two auxiliary components complete the pipeline: Context Preparation, which structures and enriches the input code for analysis, and Comment Aggregation, which aggregates similar feedback to prevent information overload.
\paragraph{Context Preparation}
This phase implements a systematic approach to prepare the input code and construct appropriate prompts for analysis. The process involves three key steps: (1) We partition each code diff based on header hunks, creating several units for analysis to manage context length and prevent excessive token consumption in subsequent processing steps. (2) We expand each segmented code block to include complete function definitions, following a controlled strategy that extends to function boundaries within four times the original diff size, or limits context to three times the original size otherwise. We utilize tree-sitter\footnote{\url{https://tree-sitter.github.io/tree-sitter/}} for precise code block identification and function boundary detection. (3) We implement a detailed change annotation system that marks each line with its status and position: old version lines are marked as ``[deleted or pre-modified @line\_number in old code]'' or [unchanged]'', while new version lines are marked as ``[added or post-modified @line\_number in new code]'' or ``[unchanged]''. These annotations preserve the original structure while providing crucial context about code modifications. 
When these contexts are prepared, we will apply them to the subsequent prompt construction. This systematic preparation ensures the model receives both comprehensive context and clear instructions, enabling more accurate and relevant review comments.

\paragraph{RuleChecker}
The RuleChecker component identifies issues using a fine-tuned LLM that integrates ByteDance's internal code standards with our taxonomy of review rules. This model examines the target code thoroughly and provides detected issues along with modification suggestions, the outputs such as in Figure ~\ref{fig:code-review-example}. 
Additionally, the component includes a rule category blocker that enables dynamic rule exclusion without model retraining when we deem them unacceptable to the user.

\paragraph{ReviewFilter} \label{par:filter}
The ReviewFilter takes a review comment as input and outputs a binary decision (yes/no) to determine whether the comment should be retained. 
This component enhances pipeline precision by addressing hallucination and factual errors in the output of RuleChecker. In practice, we reveal that RuleChecker frequently encounters hallucination-related and factual error-related bad cases when identifying issues. However, our experiments find that conventional Supervised Fine-Tuning methods for RuleChecker prove insufficient for error mitigation, even with optimized training samples and reinforcement learning approaches.
Through multiple iterations, we eventually opted to introduce the ReviewFilter component to perform a secondary validation on the results generated by RuleChecker, thereby improving the precision of the comments. The ReviewFilter component is also based on a fine-tuned LLM. 
To enable effective comment validation, we explore three distinct reasoning patterns in the model's output structure: (1) Direct Conclusion, which generates a single decision (Yes or No) token without reasoning; (2) Reasoning-First, which provides complete reasoning before concluding, necessitating full token generation like chains of thought (CoT) \cite{wei2022chain}; and (3) Conclusion-First, which outputs a decision token followed by supporting rationale, requiring only the first token for evaluation while preserving the complete output. Each pattern offers different trade-offs between precision, inference speed, and explainability, which we evaluate extensively in Section~\ref{par:cot}. 
We finally chose the Conclusion-First pattern to organize the training data. 

\paragraph{Comment Aggregation}
Multiple files and header hunks within a single MR may generate lots of similar comments, potentially overwhelming developers with redundant feedback. 
The comment Aggregation module is at the end of the pipeline. This module employs cosine similarity to determine whether comments are similar and randomly retains one comment from each similar group. Specifically, issue categories and comments are vectorized using our internal embedding model, Doubbao-embedding-large, which operates in 512 dimensions.

After going through the above pipeline, we will obtain a batch of accurate comments for each merge request. These comments will include detailed problem descriptions, modification suggestions, and will be sufficiently aggregated to prevent disruption. 

\subsection{Evaluation Metrics}
\paragraph{Prioritizing Precision in Code Review Practice.} 
In the early stages of code review development, we identify two critical insights through analysis and user studies. First, similar to alert fatigue, developers tend to ignore review comments entirely when faced with numerous comments, as they are unwilling to invest time in careful discrimination. Second, inaccurate comments in the early stages significantly damage user trust, hindering continuous iteration and improvement of the system. 
These practical challenges lead us to prioritize precision over recall in our approach.  
Regarding recall, a fundamental challenge lies in the substantial human effort required for comprehensive issue detection. Many defects only manifest after extended periods of production deployment or a large amount of manpower investment, and even well-tested code may harbour latent issues that emerge only through specific usage patterns or edge cases. 
Formally, let $C_{correct}$ represent the set of correct comments and $C_{total}$ represent all comments generated by \method{}, we define precision as: 
\begin{equation}
\text{Precision} = \frac{|C_{correct}|}{|C_{total}|} \times 100\%
\end{equation}
\paragraph{Outdated Rate for Automated Evaluation.} 
While precision serves as a crucial metric for evaluating model performance, precision measurements alone present fundamental limitations. A primary limitation is that precision cannot reflect whether developers actually accept and act upon the review comments. Additionally, manual precision assessment requires significant effort, making it challenging to conduct both large-scale evaluations and sustained monitoring over time. 
To address these limitations, particularly the need to measure developer response to comments, we introduce the \textit{"Outdated Rate"} metric, which tracks the percentage of code lines modified after being flagged by \method{}. 
Specifically, this metric is computed as:
\begin{equation}
\text{Outdated Rate} = \frac{|\{c \in C_{seen} \land  \text{isOutdated}(c)\}|}{|C_{seen}|} \times 100\%
\end{equation}
where $C_{seen}$ represents the set of comments reviewed by code committers within a one-week measurement window, and the function \text{isOutdated}($c$) returns true only if a comment $c$ is considered outdated, which occurs when if any line within its flagged code range is modified in subsequent commits. This automated measurement approach enables systematic evaluation at a scale where manual assessment becomes impractical. While the Outdated Rate doesn't definitively prove that changes were made in direct response to \method{}'s comments, it helps improve systems continuously in large-scale deployments.
\subsection{Data Flywheel: Continuous Evolution}
The effectiveness of \method{} relies on high-quality datasets and continuous feedback-driven iteration. Our data flywheel approach systematically constructs, maintains, and enhances datasets through targeted optimizations based on user feedback as illustrated in the lower part of Figure~\ref{fig:cr}.
\subsubsection{Mining Review Rules integrate Code Style Guide}
The foundation of \method{} is a comprehensive taxonomy of code review rules that integrates code style guides with practical review experience (displayed in Table~\ref{tab:code-review-taxonomy} and Section~\ref{Taxonomy}). 
Code style guides provide standardized conventions for code formatting, best practices, programming principles, and so on.
Based on the code style guide, we derive rules from two primary sources: 
\begin{itemize}
\item \textbf{Internal Static Analysis Rules}: ByteDance employs a comprehensive set of internal rules for static analysis. 
Each static analysis rule is marked with the recommendation index and acceptance rate. We select those rules with high recommendation index and high developer acceptance rate to form our review rule set. 
\item \textbf{Manual Review Comments}: Considering the deficiencies of Static Analysis Rules in understanding code semantics, we categorize internal manual review comments to extract review rules not covered by static analysis rules. This includes review rules such as ``spelling errors'', ``duplicate code'', and ``unclear code comments''.
\end{itemize}
Our current system encompasses 219 review rules across five programming languages, dynamically updated through the feedback loop described below.

\subsubsection{Dataset Construction for Model Training}
The data flywheel transforms review rules into high-quality training datasets through a systematic process:
\begin{itemize}
\item \textbf{Original Data Collection}: We extracted $120000$ code review comments from our internal code repository's MR comments, encompassing both static analysis results and manual review feedback. This original dataset served as the foundation for the primary source for subsequent data refinement.

\item \textbf{Data Refinement}: We develope a refinement process utilizing the \texttt{Doubao-Pro-32K-0828} LLM. For manual review comments, the process first filters non-substantive content (e.g., conversational elements, emojis) and classifies the remaining samples to retain rule-compliant comments. 
Then enhance these comments by expanding concise feedback into comprehensive issue descriptions with specific modification suggestions. 
For static analysis results, we refined the comments based on the specific definitions of static analysis rules while incorporating targeted modification suggestions.

\item \textbf{Quality Assurance}: We ensure dataset quality through systematic manual sampling and annotating.
To save labour costs, we employ deterministic sampling rules. 
After annotating samples from each review rule, we adjust the sampling rates—reducing the rate for high-precision review rules and increasing the rate for low-precision ones. 
\end{itemize}
This process generates refined training datasets comprising approximately 18,000 samples each for Go and front-end languages, and 5,000 samples each for other programming languages.

\subsubsection{Online Feedback and Continuous Evolution}
\begin{figure}[t!b]
\centering
\adjustbox{frame}{\includegraphics[width=1\linewidth]{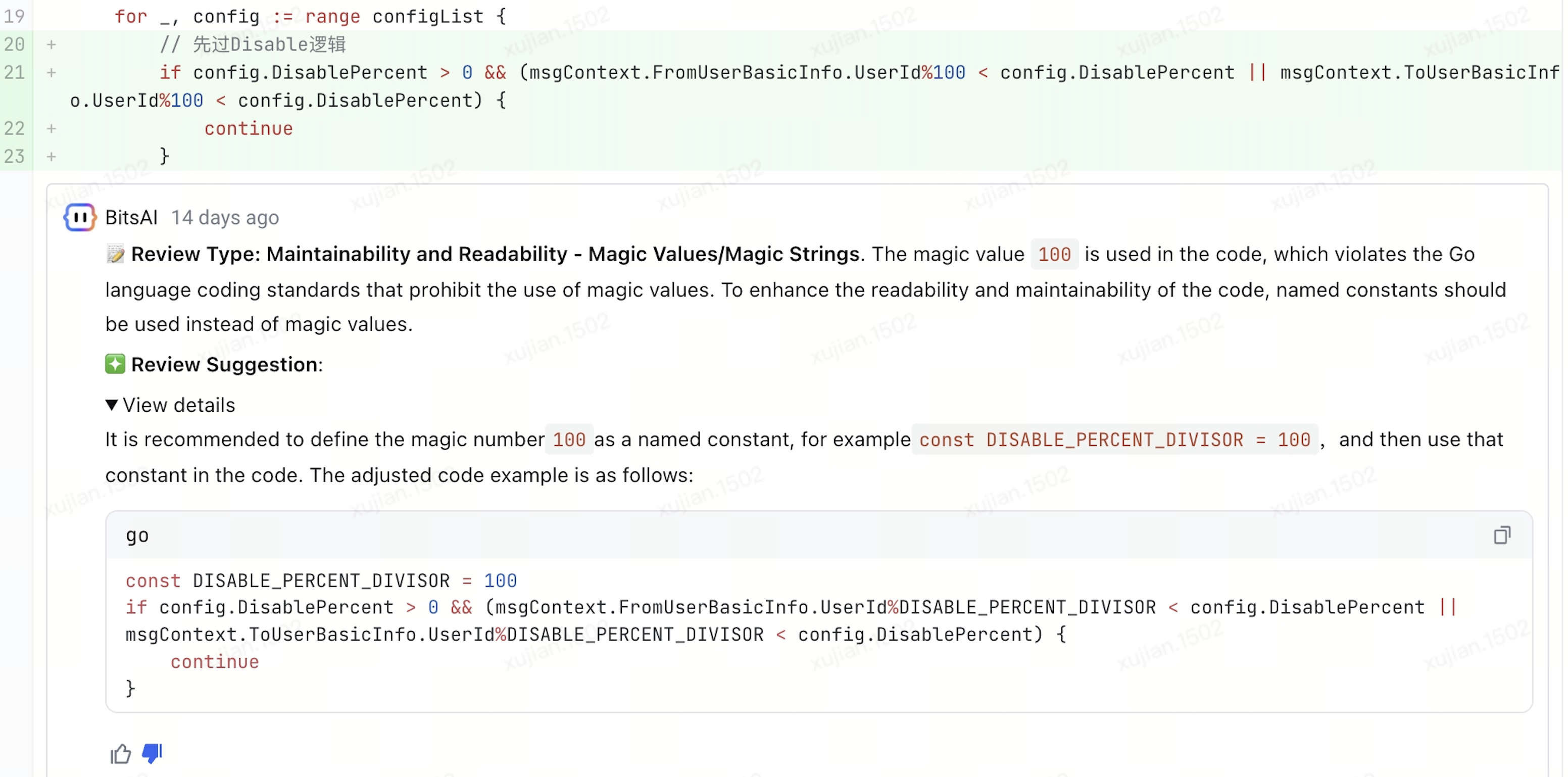}}
\caption{A Correct but Superfluous Comment}
\label{fig:case_non}
\end{figure}

After the deployment of \method{}, we conduct detailed evaluations of its online performance every week and make targeted adjustments based on the evaluation results. Our assessments primarily come from the following three aspects:
\begin{itemize}
\item \textbf{User Feedback Collection}: The most direct feedback comes from the user's likes and dislikes, often accompanied by reasons for the ratings. By analyzing the like-dislike data, we can specifically optimize the underperforming review rules. However, this type of feedback data is relatively sparse and is generally used as supplementary analysis.
\item \textbf{Manual Precision Annotation}: Each day, we sample and annotate the online data (sampling generally does not exceed 10\%). The annotation results are compiled weekly, providing an accurate measurement of the review rules' precision. For review rules with low precision, we will collect the corresponding samples to retrain the LLM. The training data for the aforementioned RuleFiler component mainly comes from this source.
\item \textbf{Outdated Rate Monitoring}: Every week, we track changes in the Outdated Rate of each review rule. For scenarios where the precision is high but the Outdated Rate is consistently low, we consider whether the review rule is not being accepted by users and may decide on decommissioning it.
\end{itemize}
These diverse feedback channels provide valuable insights into both the technical precision and practical utility of our review rules. For example, Figure~\ref{fig:case_non} demonstrates this with an example where \method{} flags the use of a magic number ``100''. While this suggestion aligns with Go language coding standards that discourage magic numbers, the user responds with a dislike. Such cases that users dislike highlight the need for a more comprehensive evaluation framework beyond simple precision metrics.

To address this challenge, we develop a systematic approach using both precision and Outdated Rates as key metrics for evaluating review rules. Rules are assessed based on specific criteria: an Outdated Rate of around 25\% (±5\%) with a precision of around 65\% (±5\%) for 14 days. This dual-metric evaluation framework enables us to make data-driven decisions about retaining or removing specific review rules.
Our comprehensive taxonomy of code review rules as shown in Table ~\ref{tab:code-review-taxonomy} serves as the foundation for this evolution process. By mapping all online feedback samples to specific rules within this taxonomy, we can efficiently collect data and conduct targeted evaluations. This systematic approach creates a continuous feedback loop: user interactions provide data for rule evaluation, evaluation results guide rule refinements, and refined rules generate new feedback. Through this iterative process, we maintain a dynamic review system that consistently adapts to user needs while maintaining high-quality standards.

\subsection{Implementation of \method{}}
\begin{figure}[t!b]
    \centering
    \adjustbox{frame}{\includegraphics[width=1\linewidth]{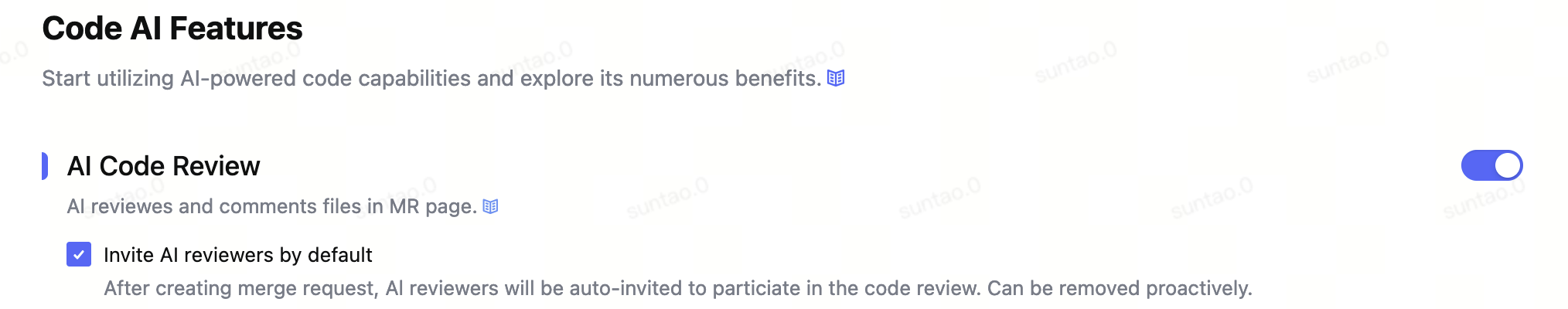}}
    \caption{\method{} settings interface for enabling review participation and default reviewer invitation.}
    \label{fig:codeai}
\end{figure}

\begin{figure}[t!b]
\centering
\adjustbox{frame}{\includegraphics[width=\linewidth]{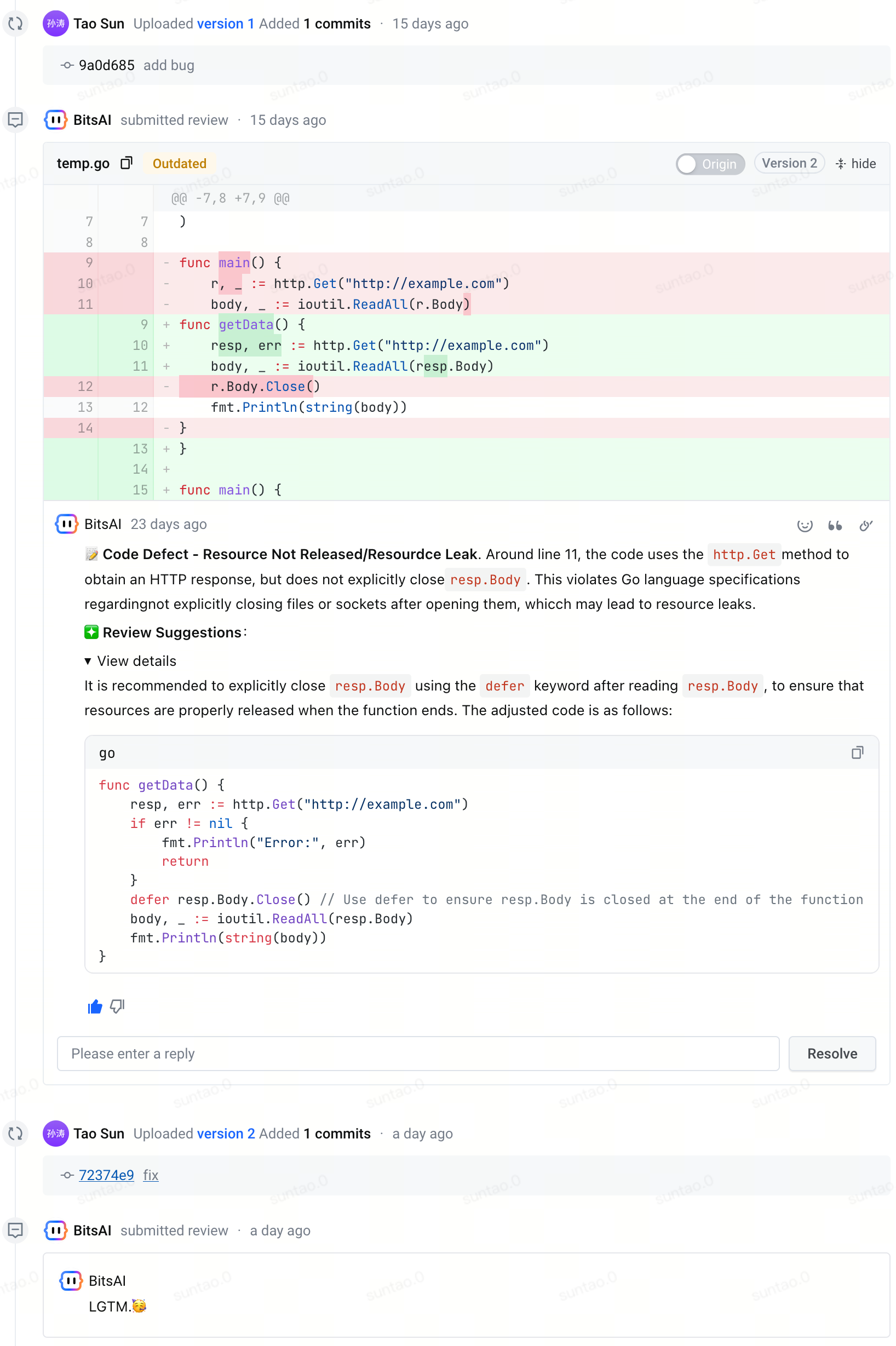}}
\caption{The MR interface shows the ``outdated'' status of the initial review comments and \method{}'s final approval ("LGTM" - Looks Good To Me), indicating that the developer successfully resolves issues through next commit.} \label{fig:lgtm}
   \end{figure}

Developers can easily enable \method{} features as shown in Figure~\ref{fig:codeai}. The \textbf{input} to our system is a code diff, and the \textbf{output} is a comprehensive review comment as shown in Figure~\ref{fig:code-review-example}. As illustrated in Figure~\ref{fig:lgtm}, \method{} automatically identifies potential issues in the code waiting to be merged, confirms the review category, pinpoints problematic code lines, and provides relevant comments. After developers address these review comments and make the necessary code modifications, \method{} re-evaluates the changes and, as shown in Figure~\ref{fig:lgtm}, marks the original review comments as ``outdated'' and provides an ``LGTM'' (Looks Good To Me) approval, indicating that the code modifications successfully addressed the identified issues and now meets the required quality standards.

%% file: fig/prompt/task.tex
\newtcolorbox[auto counter, number within=section]{mybox}[2][]{%
  colback=white,
  colframe=black,
  width=\linewidth,
  arc=0mm,
  boxrule=0.1mm,
  fonttitle=\small,
  fontupper=\footnotesize,
  before upper={\raggedright},
  #1
}
\begin{figure}[t]
\centering
\begin{mybox}
{\fontsize{10}{12}\selectfont\emph{\textbf{Input:}}}
\vspace{0.5em}
\begin{verbatim}
diff --git a/calculate.go b/calculate.go
index a123456..b789012 100644
--- a/calculate.go
+++ b/calculate.go
@@ -7,3 +7,5 @@ func CalculateArea(radius float64) float64 {
-    const pi = 3.14
+    const pi = 3.14159
+    area := pi * radious * radious
+    return area
 }
\end{verbatim}
\vspace{-0.5em}
\textcolor{gray}{\hrule}
\vspace{1em}
{\fontsize{10}{12}\selectfont\emph{\textbf{Output:}}}
\vspace{0.5em}
\begin{Verbatim}[breaklines=true,commandchars=\|\{\}]
|textbf{Line}: 9
|textbf{Category}: Spelling Error
|textbf{Severity}: Medium
|textbf{Issue}: In the 9th line of the modified code, the variable name 'radious' has a spelling error.
|textbf{Suggestion}: Please correct the spelling error in the variable name (change 'radious' to 'radius').
\end{Verbatim}
\end{mybox}
\caption{An Example of Input and Output in \method{}}
\label{fig:code-review-example}
\end{figure}

%% file: sections/04Experiments.tex
\section{Experiments and Analysis}

This section presents a comprehensive evaluation of \method{} through both offline experiments and large-scale industrial deployment. We first detail our model training approach and configuration, followed by a rigorous offline evaluation of code review capabilities using a carefully curated dataset. We then conduct an ablation study to validate the necessity of ReviewFilter. The analysis extends to online performance metrics in production environments, including precision trends and user retention rates. Throughout these experiments, we demonstrate \method{}'s effectiveness across multiple dimensions of code review, from technical precision to practical utility in real-world development scenarios.

\subsection{Model Training}

The development of \method{} is primarily driven by enterprise security requirements and data privacy considerations. As a result, we utilize \texttt{Doubao-Pro-32K-0828}, ByteDance's developed LLM, which ensures compliance with our security policies while maintaining high-performance standards. It's one of the most advanced LLMs and has a rich knowledge base and robust analytical capabilities.

We employ a fine-tuning approach using the Low-Rank Adaptation (LoRA)\cite{hu2021lora} technique on the \texttt{Doubao-Pro-32K-0828} for both RuleChecker and ReviewFilter. Based on our analysis of historical review data, which shows that 99\% of review samples contain fewer than 8192 tokens, we use \texttt{Doubao-Pro-32K-0828} with an 8192 sequence length. We fine-tune it for 5 epochs with a batch size of 8. The LoRA configuration includes a rank of 128 and an alpha of 256. We use a learning rate of 0.00005 with gradient accumulation over 1 step and a warmup step rate of 0.05 to stabilize early training.

\subsection{Evaluation of Code Review Capabilities}

To evaluate \method{}'s effectiveness in code review, we collect an offline dataset consisting of 1397 cases sampled from the production codebase, where 767 samples violate and 630 samples follow the code best practices. These cases are drawn from the taxonomy of review rules, as categorized in Table~\ref{tab:code-review-taxonomy}.

Following the LLM-as-a-judge methodology\cite{zheng2023judging}, we employ \texttt{Doubao-Pro-32K-0828} to evaluate automatically our business code dataset while preserving data confidentiality. 
The review comment is deemed correct only if the model determines it aligns with the ground truth. 
The experimental results are presented in Table~\ref{tab:complex-review-challenges}, where we compare our approach against strong baseline models including Qwen2.5-Coder-32b-instruct\cite{qwen2coder} and Deepseek-v2.5\cite{deepseekV2}. We evaluated two versions: (1) a base version (\method{} w/o Taxonomy) trained on randomly sampled human internal review data without taxonomic classification (similar to existing open-source approaches\cite{liAutomatingCodeReview2022}), and (2) a taxonomy-guided version (\method{}) where the training data is specifically constructed according to our taxonomy of review rules while maintaining the same data volume. The results demonstrate that while our base \method{} already outperforms baseline models, the taxonomy-guided version achieves substantially higher precision across all categories, reaching 57.03\% overall precision compared to the base model's 16.83\%. These results demonstrate that our taxonomy enhances review precision.

\begin{table}[h]
\resizebox{\linewidth}{!}{
\begin{tabular}{@{}cccccc@{}}
\toprule
\multirow{2}{*}{Model} &
  \multirow{2}{*}{\begin{tabular}[c]{@{}c@{}} Review \\ Dimension\end{tabular}} &
  \multicolumn{2}{c}{Only RuleChecker} &
  \multicolumn{2}{c}{With ReviewFilter} \\ \cmidrule(l){3-6} 
 &
   &
  \begin{tabular}[c]{@{}c@{}}Precision \\ (\%)\end{tabular} &
  \begin{tabular}[c]{@{}c@{}}Recall \\ (\%)\end{tabular} &
  \begin{tabular}[c]{@{}c@{}}Precision\\  (\%)\end{tabular} &
  \begin{tabular}[c]{@{}c@{}}Recall \\ (\%)\end{tabular} \\ \midrule
Qwen2.5-Coder-32b-instruct &
  \multirow{5}{*}{ALL} &
  10.14 &
  21.90 &
  10.62 &
  20.86 \\
Deepseek-v2.5 &
   &
  9.27 &
  16.30 &
  9.50 &
  16.17 \\
Doubao-Pro-32K-0828 &
   &
  7.65 &
  13.56 &
  7.70 &
  13.56 \\
\method \ w/o Taxonomy &
   &
  16.83 &
  31.55 &
  30.92 &
  22.29 \\
\method &
   &
  \textbf{57.03} &
  45.50 &
  \textbf{65.59} &
  39.77 \\ \midrule
Qwen2.5-Coder-32b-instruct &
  \multirow{5}{*}{\begin{tabular}[c]{@{}c@{}}Security \\ Vulnerability\end{tabular}} &
  18.52 &
  35.71 &
  20.83 &
  35.71 \\
Deepseek-v2.5 &
   &
  14.29 &
  23.81 &
  14.49 &
  23.81 \\
Doubao-Pro-32K-0828 &
   &
  9.86 &
  16.67 &
  10.00 &
  16.67 \\
\method \ w/o Taxonomy &
   &
  10.00 &
  16.67 &
  19.23 &
  11.90 \\
\method &
   &
  \textbf{58.82} &
  47.62 &
  \textbf{61.29} &
  45.24 \\ \midrule
Qwen2.5-Coder-32b-instruct &
  \multirow{5}{*}{\begin{tabular}[c]{@{}c@{}}Code \\ Defect\end{tabular}} &
  7.62 &
  16.89 &
  7.92 &
  16.23 \\
Deepseek-v2.5 &
   &
  8.61 &
  16.23 &
  8.96 &
  16.23 \\
Doubao-Pro-32K-0828 &
   &
  4.72 &
  8.94 &
  4.75 &
  8.94 \\
\method \ w/o Taxonomy &
   &
  16.20 &
  34.44 &
  26.19 &
  21.85 \\
\method &
   &
  \textbf{53.88} &
  43.71 &
  \textbf{68.92} &
  33.77 \\ \midrule
Qwen2.5-Coder-32b-instruct &
  \multirow{5}{*}{\begin{tabular}[c]{@{}c@{}}Maintainability \\ and Readability\end{tabular}} &
  11.06 &
  24.54 &
  11.56 &
  22.98 \\
Deepseek-v2.5 &
   &
  8.97 &
  15.40 &
  9.11 &
  15.14 \\
Doubao-Pro-32K-0828 &
   &
  9.92 &
  17.23 &
  9.97 &
  17.23 \\
\method \ w/o Taxonomy &
   &
  18.15 &
  31.85 &
  35.94 &
  24.02 \\
\method &
   &
  \textbf{58.12} &
  46.74 &
  \textbf{63.60} &
  43.34 \\ \midrule
Qwen2.5-Coder-32b-instruct &
  \multirow{5}{*}{\begin{tabular}[c]{@{}c@{}}Performance \\ Issue\end{tabular}} &
  14.29 &
  20.00 &
  14.81 &
  20.00 \\
Deepseek-v2.5 &
   &
  13.46 &
  17.50 &
  13.46 &
  17.50 \\
Doubao-Pro-32K-0828 &
   &
  7.69 &
  10.00 &
  7.84 &
  10.00 \\
\method \ w/o Taxonomy &
   &
  16.67 &
  22.50 &
  42.11 &
  20.00 \\
\method &
   &
  \textbf{72.00} &
  45.00 &
  \textbf{72.00} &
  45.00 \\ \bottomrule
\end{tabular}
}
\caption{Performance on Code Review Evalution}
\label{tab:complex-review-challenges}
\end{table}

\paragraph{Ablation Study for ReviewFilter.} 
To evaluate the necessity of the ReviewFilter component, we conduct an ablation study comparing model performance with and without this filtering stage, as shown in Table~\ref{tab:complex-review-challenges}. Our experiments reveal that advanced LLMs' raw outputs, even with extensive fine-tuning, fail to meet the precision requirements for production deployment. This limitation manifests in persistent hallucinations across various code review scenarios. For example, we observed that the fine-tuned RuleCheck module often incorrectly identifies formatting issues. In one case, the output of message is like: ``The variable name `basicInfoInstantHome' does not comply with coding standards, the underscore `\_' should be removed, changing it to `basicInfoInstantHome'.'' This message is evidently a hallucination since the variable name `basicInfoInstantHome' already complies with the coding standards and does not contain any underscores. In another case, it mistakes the function name `WhenAwemeStartFrom2580' as containing a magic number, demonstrating how fine-tuned LLMs can misinterpret code context without proper filtering. The ablation results show that the addition of ReviewFilter consistently improves precision across all categories - increasing overall precision from 54.50\% to 67.12\% in the taxonomy-guided model. These quantitative improvements, combined with the qualitative examples of prevented hallucinations, demonstrate that the ReviewFilter is an essential component for achieving production-grade reliability in automated code review.

\begin{table}[t!b]
\resizebox{\linewidth}{!}{
\begin{tabular}{@{}ccccc@{}}
\toprule
\textbf{Reasoning Pattern} &
  \textbf{\begin{tabular}[c]{@{}c@{}}Precision \\ (\%)\end{tabular}} &
  \textbf{\begin{tabular}[c]{@{}c@{}}Recall\\ (\%)\end{tabular}} &
  \textbf{\begin{tabular}[c]{@{}c@{}}Filter Rate\\ (\%)\end{tabular}} &
  \textbf{\begin{tabular}[c]{@{}c@{}}Inference Time\\ (s/sample)\end{tabular}} \\ \midrule
Direct Conclusion    & 63.27          & 77.50          & 38.75          & \textbf{1.7} \\
Reasoning-First  & 65.80          & \textbf{81.80} & 30.00          & 31.0           \\
Conclusion-First & \textbf{77.09} & 69.00          & \textbf{55.25} & \textbf{1.7} \\ \bottomrule
\end{tabular}
}
\caption{Performance Comparison of Reasoning Patterns}
\label{tab:cot-patterns}
\end{table}

\paragraph{Reasoning Patterns of ReviewFilter.} \label{par:cot}
To optimize our comment validation mechanism in ReviewFilter, we evaluate three reasoning patterns described in Section~\ref{par:filter}. We collect an additional 400 related data for the experiment. As indicated in Table~\ref{tab:cot-patterns}, the Direct Conclusion pattern records the lowest precision. Significantly, enriching the training dataset with a reasoning process can enhance the performance of the ReviewFilter. A comparative analysis of the reasoning patterns reveals that the Reasoning-First pattern attains a superior recall of 81.80\%, albeit at the cost of a noticeably prolonged inference time of 31s/sample, rendering it less feasible for deployment in production scenarios. Conversely, the Conclusion-First pattern, while obtaining the lowest recall rate at 69.00\% and the highest filter rate, measured as the ratio of comments identified as erroneous by ReviewFilter, of 55.25\%. This pattern notably improves the precision to 77.09\%. Furthermore, the inference time for the Conclusion-First pattern is comparably short, akin to that of the Direct Conclusion pattern, due to its dependence predominantly on the initial token generation during inference. Given the prioritization of precision over recall in code review contexts, the Conclusion-First pattern offers a balance between effectiveness and operational efficiency, thereby making it the preferred choice.

\subsection{Online Product Performance}

To illustrate the effectiveness of our data flywheel approach, we track the performance of \method{} over time.

\begin{figure}[tb]
    \centering
    \includegraphics[width=1\linewidth]{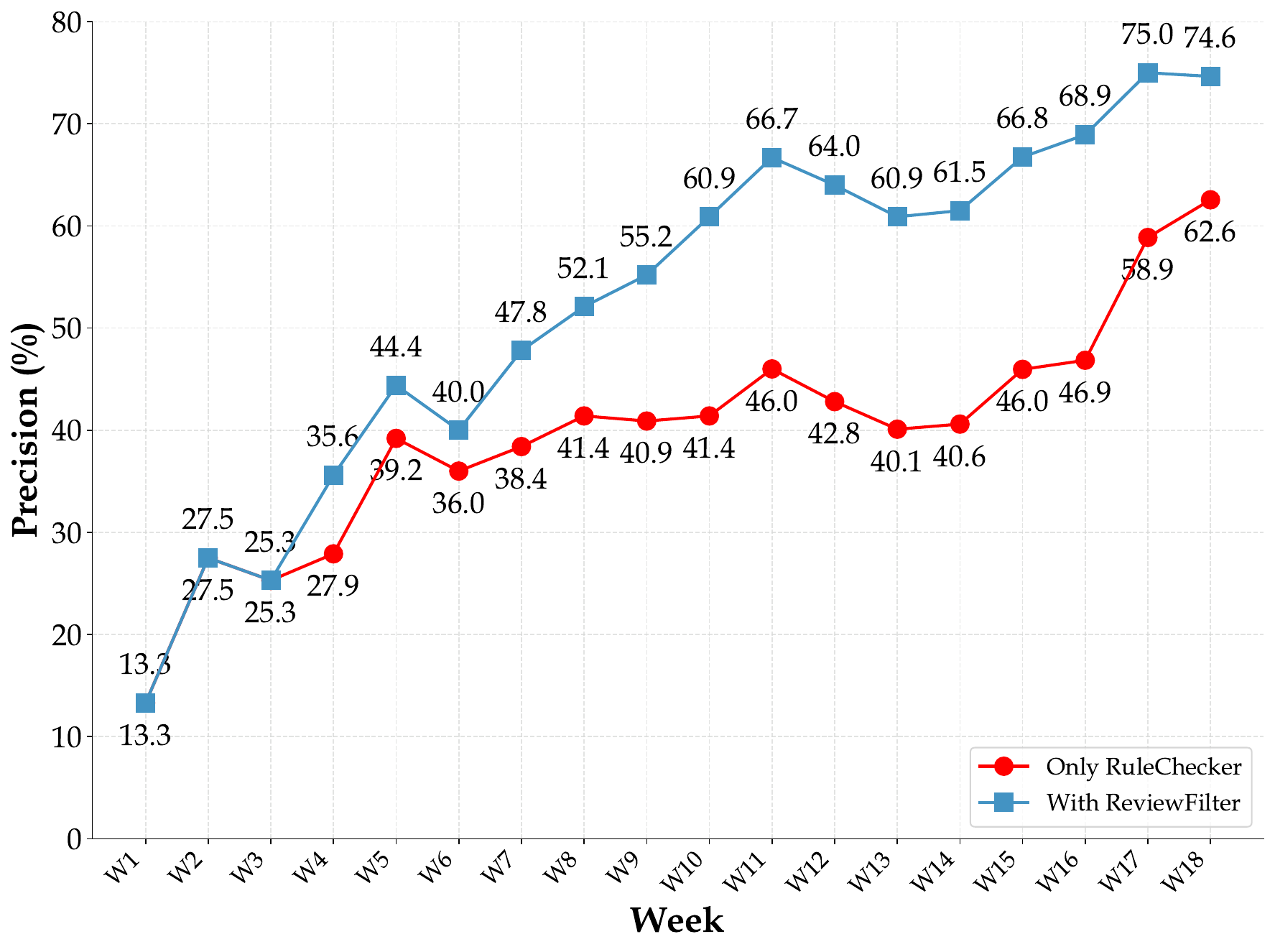}
    \caption{Weekly Progression of \method{} Precision}
    \label{fig:model-performance-over-time}
\end{figure}

\paragraph{Precision Trends.} 
Figure~\ref{fig:model-performance-over-time} illustrates the precision trends of \method{} over an 18-week period. In the initial three weeks, without the taxonomy of review rules method and two-stage approach, the overall precision remained stagnant at around 25\% (with no distinction between RuleChecker and ReviewFilter at this stage). After implementing the taxonomy of review rules method combined with the two-stage approach, both components began to show gradual improvement. The RuleChecker's precision increased from an initial 27.9\% to 62.6\%, while the ReviewFilter's precision rose from 35.6\% to a peak of 75\%. These improvements demonstrate the sustainable iterative nature of our approach. Notably, figure~\ref{fig:model-performance-over-time} shows that RuleChecker's precision consistently remains lower than ReviewFilter's precision, enabling the ReviewFilter to deliver superior results to users and thereby validating its essential role.

\begin{figure}[tb]
    \centering
    \includegraphics[width=1\linewidth]{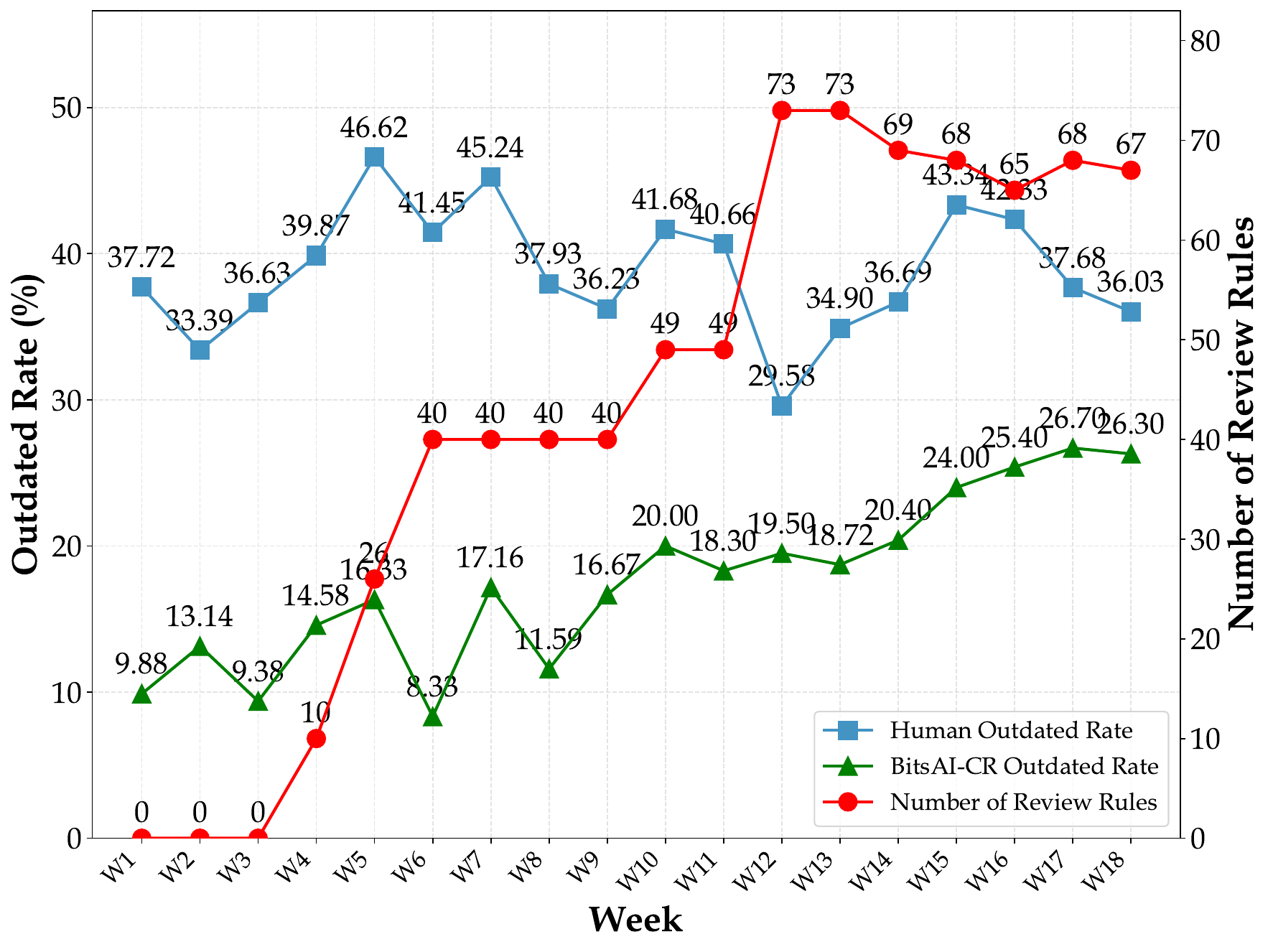}
    \caption{Weekly Progression of Outdated Rates and Review Rule Counts in Go Language}    \label{fig:outdated_and_Resolved_rates}
\end{figure}

\paragraph{Outdated Rate Trends.}
The effectiveness of our approach is demonstrated in Figure~\ref{fig:outdated_and_Resolved_rates}, which tracks three key metrics over an 18-week period only in Go Language. 
We select Go because it is a primary language at ByteDance, which establishes support and optimization processes from early code reviews, making it ideal for observing our methodology's effect. 
During the first 10 weeks, despite the continuous increase in review rules, the Outdated Rate remains relatively stable at around 15\%, even as the precision metrics show improvement. 
After expanding our review rules to 73, we observe a significant increase in the Outdated Rate, reaching approximately 20\%, coinciding with a peak precision rate of 63\%. Following this period, the Outdated Rate stabilizes temporarily. Starting from week 14, we begin optimizing \method{} by removing underperforming review rules based on Outdated Rate and precision metrics, which leads to a gradual increase in the Outdated Rate, ultimately reaching a peak of 26.7\% by week 18. For comparison, the Human Outdated Rate at ByteDance representing how often code flagged by human reviewers gets modified, fluctuates between 35\%–46\%, serving as a baseline for effective code review impact. The gradual convergence of \method{}'s Outdated Rate toward human-level performance demonstrates the effectiveness of our data flywheel.

\paragraph{User Feedback and Expert Interview.}
To validate the effectiveness of \method{}, we conducted a comprehensive evaluation combining quantitative surveys (N=137) and in-depth expert interviews (N=12). The survey participants were randomly selected across different programming languages, with Go (50\%), frontend technologies (25\%), and other languages including Java and Python (25\%) represented. The results showed strong positive reception, with 74.5\% (102/137) of users affirming the value and effectiveness of \method{}'s code reviews. Among the remaining 25.5\% (35/137) who suggested improvements, the feedback was categorized as follows: incorrect comments (10.9\%, 15/137), correct but unnecessary comments (12.4\%, 17/137), irrelevant feedback (1.5\%, 2/137), and comprehension difficulties (0.7\%, 1/137). To complement the survey data, we conducted structured 10-minute interviews with 12 expert developers who had used \method{} for more than one month. The participants represented diverse roles including frontend, backend, and algorithm development. Table~\ref{tab:expert-feedback} summarizes the key findings from these interviews, categorized by usage patterns, quality assessment, performance considerations, and feature requests.

\begin{table}[tb]
\centering
\caption{Analysis of Expert Interview Feedback (N=12)}
\label{tab:expert-feedback}
\resizebox{1\linewidth}{!}{
\begin{tabular}{@{}cc@{}}
\toprule
\textbf{Aspect} & \textbf{Key Findings} \\
\midrule
\multirow{3}{*}{\textbf{\begin{tabular}[c]{@{}c@{}}Quality\\ Assessment\end{tabular}}} & Unanimous agreement on \textbackslash{}method\{\} utility (12/12) \\
 & Quality improvement suggestions (3/12) \\
 & The Go feels good, other languages need to be improved (4/12) \\ \midrule
\multirow{2}{*}{\textbf{\begin{tabular}[c]{@{}c@{}}Performance\\ Concerns\end{tabular}}} & Feedback on latency of review generation (7/12) \\
 & Strong desire for improved processing speed (9/12) \\ \midrule
\multirow{5}{*}{\textbf{\begin{tabular}[c]{@{}c@{}}Feature\\ Requests\end{tabular}}} & Support for customizable review rules (11/12) \\
 & Don't review the code auto-generated by the framework (2/12) \\
 & More program languages review support (2/12) \\
 & One-click suggestion application functionality (11/12) \\
 & SDK provision for pre-push review integration (1/12) \\ \bottomrule
\end{tabular}
}
\end{table}

\subsection{Large-Scale Industrial Deployment}

Before full-scale implementation, we conduct canary testing to validate the system's performance and user acceptance. This phased deployment approach ensures minimal disruption to users and also allows us to rapidly collect user feedback data and make immediate adjustments based on online responses.

\begin{figure}[t]
    \centering
    \includegraphics[width=1\linewidth]{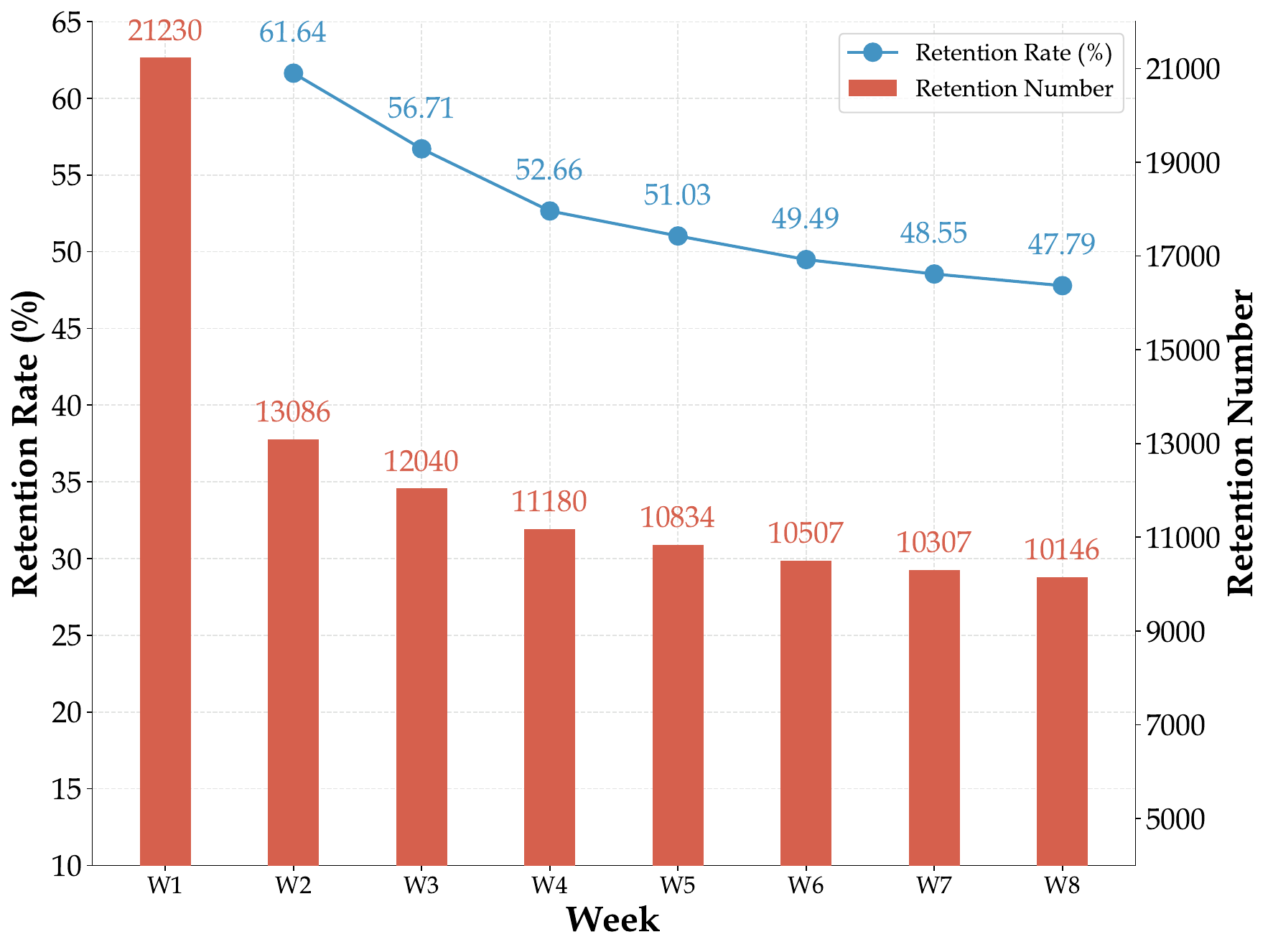}
    \caption{Weekly User Retention Rate of \method{}}
    \label{fig:user_retention_rate}
\end{figure}

Currently,\method{} fully deploys across ByteDance's development teams. It boasts over \textbf{12k Weekly Active Users (WAU)} and \textbf{210k Weekly Page Views (WPV)}, demonstrating its widespread adoption and effective integration into the development workflow.
Furthermore, as shown in Figure~\ref{fig:user_retention_rate}, the system shows a second-week retention rate of 61.64\%, and can still maintain a high retention rate of around 48\% through the eight weeks. 
To our knowledge, this represents the first documented retention rate benchmark for large-scale code intelligence tools, providing valuable reference metrics for other organizations implementing similar systems.

%% file: sections/05Results.tex
\section{Lessons Learned and Practical Insights}
The large-scale implementation of \method{} has yielded valuable insights that can guide other organizations in developing and deploying automated code review systems. We present our key findings across the following critical dimensions:

\find{\textbf{Taxonomy of Review Rules} \faIcon{hand-point-right} \textit{enables systematic code issue categorization, data collection, and performance evaluation.}} 
Our experience demonstrates that a well-defined taxonomy of code review rules serves as the cornerstone for building effective automated review systems. The taxonomy (Table~\ref{tab:code-review-taxonomy}) provides crucial benefits: (1) it enables systematic issue categorization and detection, guiding the development of \method{} which improved precision from 30.92\% to 65.59\% with ReviewFilter in the Go language, (2) it facilitates structured data collection and labeling, ensuring consistent training data quality, and (3) it provides clear criteria for evaluating and improving system performance across different review dimensions. Unlike traditional approaches that lack systematic data-driven evolution, our taxonomy-driven approach enables more precise and actionable code review through structured issue classification and targeted optimization.

\find{\textbf{Two-Stage Review Generation} \faIcon{hand-point-right} \textit{enhances automated code review reliability by validating identified issues.}}
Our findings demonstrate that a two-stage review generation approach, combining RuleChecker and ReviewFilter, is crucial for achieving production-grade reliability in automated code review. 
While LLMs can identify potential issues, their tendency to produce false positives and hallucinations necessitates a robust validation mechanism. Our implementation of ReviewFilter as a dedicated validation module significantly improved the \method{}'s reliability, increasing overall precision from 60\% to 75\%. This improvement demonstrates that a dedicated filter module is a critical component for building dependable automated code review systems.

\find{\textbf{Precision and Outdated Rate Metrics} \faIcon{hand-point-right} \textit{guides data flywheel optimization through user-centric evaluation.}}
Our deployment track record demonstrates that prioritizing precision over recall metrics is crucial for successful automated code review adoption. High precision builds user trust early without disturbing users or causing them to abandon utilizing \method{}, encouraging continued system utilization and enabling iterative improvements. However, we discover that traditional precision metrics alone are insufficient, as they require unsustainable manual labeling and fail to capture user acceptance of review rules. To address these limitations, we introduce the Outdated Rate metric as a complementary measure that reflects user acceptance of review suggestions. The combination of precision and Outdated Rate metrics proves instrumental in driving our data flywheel, facilitating large-scale deployment, and achieving positive user feedback through continuous system refinement.

%% file: sections/06RelatedWork.tex
\section{Related Work}
\paragraph{Code Large Language Models}
Code LLMs~\cite{code_bert, huang2024opencoder,zheng2024opencodeinterpreter} represent a crucial vertical within the broader LLM domain. These models advance software engineering by enhancing code intelligence\cite{zhang2023unifying}. Recent models, including GPT-4\cite{gpt4}, Llama3\cite{llama31}, Qwen2.5-Coder\cite{qwen2coder}, and DeepSeek-V2.5\cite{deepseekV2}, demonstrate significant capabilities in multilingual code generation and debugging tasks. Code LLMs transform various aspects of software development, spanning code generation~\cite{sun2024unicoder, chai2024mceval, wang2024exploring, wang2024systematic}, program repair~\cite{sunrepofixeval, liu2024mdeval}, log parsing~\cite{li2024exploring}, web design~\cite{xiao2024interaction2code, wan2024automatically}, and other applications~\cite{liu2024fullstack, yang2024evaluating}.
\paragraph{LLM-Based Code Review}
LLM applications in code review represent an active research area that encompasses both implementation and evaluation approaches. Foundational work establishes deep learning for code review, with studies like CodeReviewer~\cite{liAutomatingCodeReview2022} and related research~\cite{tufanoUsingPreTrainedModels2022, chenCodeReviewerRecommendation2022, liAUGERAutomaticallyGenerating2022, yinAutomaticCodeReview2023} exploring pre-trained models for review tasks. AutoCommenter~\cite{vijayvergiyaAIAssistedAssessmentCoding2024} advances this field by analysing code compliance with best practices, though it limits output to Google's best practices URLs without specific code suggestions. LLaMA-Reviewer~\cite{luLLaMAReviewerAdvancingCode2023} marks a significant advance through LLAMA model fine-tuning for code review. Recent research investigates hyperparameters, fine-tuning strategies, and prompt engineering~\cite{pornprasitFineTuningPromptEngineering2024, guoExploringPotentialChatGPT2024, fanExploringCapabilitiesLLMs2024, yuFinetuningLargeLanguage2024}, while studies on multi-agent systems~\cite{tangCodeAgentAutonomousCommunicative2024, rasheedAIpoweredCodeReview2024} expand automated review capabilities. Parallel evaluation efforts include EvaCRC~\cite{yangEvaCRCEvaluatingCode2023}, which introduces a framework for assessing review comments. Additional studies~\cite{dong-kyuGPTbasedCodeReview2024, ahmedCanLLMsReplace2024, koutchemeEvaluatingLanguageModels2024, naikCRScoreGroundingAutomated2024} explore review quality assessment and human-AI consistency. Notable findings indicate LLMs can surpass human performance in error detection~\cite{mcaleeseLLMCriticsHelp2024}, while research also addresses cognitive biases in review processes~\cite{jetzenDebiasingCodeReview2024}.